\documentclass[twocolumn,showpacs,preprintnumbers,amsmath,amssymb]{revtex4}

\usepackage{slashed,bbm,nicefrac}

\newcommand{\re}{\mathop{\mathrm{Re}}\nolimits}
\newcommand{\im}{\mathop{\mathrm{Im}}\nolimits}
\newcommand{\adj}{\mathop{\mathrm{adj}}\nolimits}

\begin{document}

\preprint{DESY 13-094, NSF-KITP-13-122\hspace{10.15cm}
ISSN 0418--9833}
\preprint{August 2013\hspace{15.5cm}}

\title{\boldmath
All-order renormalization of propagator matrix for fermionic system with flavor
mixing}

\author{Bernd A. Kniehl}\thanks{%
Permanent address: {II.} Institut f\"ur Theoretische Physik,
Universit\"at Hamburg, Luruper Chaussee 149, 22761 Hamburg, Germany.}


\affiliation{Kavli Institute for Theoretical Physics, University of California,
Santa Barbara, CA~93106-4030, USA}

\date{\today}

\begin{abstract}
We consider a mixed system of Dirac fermions in a general parity-nonconserving
theory and renormalize the propagator matrix to all orders in the pole scheme,
in which the squares of the renormalized masses are identified with the complex
pole positions and the wave-function renormalization (WFR) matrices are
adjusted in compliance with the Lehmann-Symanzik-Zimmermann reduction
formalism.
We present closed analytic all-order expressions and their expansions through
two loops for the renormalization constants in terms of the scalar,
pseudoscalar, vector, and pseudovector parts of the unrenormalized self-energy
matrix, which is computable from the one-particle-irreducible Feynman diagrams
of the flavor transitions.
We identify residual degrees of freedom in the WFR matrices and propose an
additional renormalization condition to exhaust them.
We then explain how our results may be generalized to the case of unstable
fermions, in which we encounter the phenomenon of WFR bifurcation.
In the special case of a solitary unstable fermion, the all-order-renormalized
propagator is presented in a particularly compact form.
\end{abstract}

\maketitle

The experiments at the CERN Large Hadron Collider have radically changed the
landscape of particle physics.
In fact, a new weak neutral resonance, which very much looks like the missing
link of the Standard Model (SM), has been discovered \cite{Aad:2012tfa}, while,
despite concerted endeavors by armies of experimental and theoretical
physicists, no signal of new physics beyond the SM has emerged so far.
Within the present experimental precision, this new particle shares the spin,
parity, and charge-conjugation quantum numbers $J^{PC}=0^{++}$ and the coupling
strengths with the SM Higgs boson $H$, and its mass $(125.6\pm0.3)$~GeV lies
well inside the $M_H$ range predicted within the SM through global analyses of
electroweak (EW) precision data, and it almost perfectly coincides with
state-of-the-art determinations of the $M_H$ lower bound, $(129.6\pm1.5)$~GeV,
from the requirement that the SM vacuum be stable way up to the scale of the
Planck mass \cite{Bezrukov:2012sa}.
If the pole mass $m_t$ of the top quark, which, in want of a rigorous
determination at the quantum level, is presently identified with a Monte-Carlo
parameter \cite{Beringer:1900zz}, were just lower by an amount of the order of
its total decay width $\Gamma_t=(2.0\pm0.5)$~GeV \cite{Beringer:1900zz},
then the agreement would be perfect, implying that EW symmetry breaking is
likely to be determined by Planck-scale physics.
In a way, this would solve the longstanding hierarchy problem of the SM.
The Nobel Prize in Physics 2013 has just been awarded jointly to Englert and
Higgs for the theoretical discovery of the Higgs mechanism.

Obviously, nature is telling us that the SM is more robust and fundamental than
commonly accepted in recent years.
This provides a strong motivation for us to deepen and complete our
understanding of the field-theoretic foundations of the SM.
After all, we are dealing here with a renormalizable quantum field theory
\cite{'tHooft:1971fh}.
The Nobel Prize in Physics 1999 was awarded to 't~Hooft and Veltman
{\it for elucidating the quantum structure of EW interactions in physics}.
The on-shell renormalization of the SM was established to all orders of
perturbation theory using the algebraic method \cite{Grassi:1995wr}.
However, all particles were assumed to be stable, neutrinos were taken to be
massless, and quark flavor mixing was neglected.
To eliminate these unrealistic assumptions, one needs to develop a pole scheme
of mixing renormalization for unstable particles valid to all orders.
Apart from being conceptually desirable, this is becoming of major
phenomenological importance, even more so because mixing and instability of
elementary particles concur in nature.
This requires generalized concepts for flavor-changing propagators and
vertices.
In the SM with massless neutrinos, these are the propagator matrices of the up-
and down-type quarks and their charged-current vertices, which involve the
Cabibbo-Kobayashi-Maskawa (CKM) \cite{Cabibbo:1963yz} quark mixing matrix.
This pattern carries over to the lepton sector if the neutrinos are massive
Dirac fermions, and the analogue of the CKM matrix is the
Pontecorvo-Maki-Nakagawa-Sakata \cite{Pontecorvo:1957cp} neutrino mixing
matrix.

The renormalization of fermion mixing matrices was treated in
Ref.~\cite{Kniehl:2006bs} and the references cited therein.
As for the renormalization of propagator matrices of mixed systems of
fermions, the situation is as follows.
In Ref.~\cite{Donoghue:1979jq}, an early treatment of finite renormalization
effects both for quarks in hadronic bound states and leptons may be found.
In Ref.~\cite{Gambino:1999ai}, the ultraviolet (UV) renormalization of the
fermion masses was considered, and the pole masses were shown to be gauge
independent to all orders in the SM using Nielsen identities
\cite{Nielsen:1975fs}, both for stable and unstable fermions.
In Ref.~\cite{Kniehl:2012zb}, the UV renormalization of the fermion fields was
discussed for the case of stability, and the dressed propagator matrices were
written in closed form, both for the unrenormalized and renormalized versions.
Furthermore, it was explicitly proven that the WFR conditions proposed in
Ref.~\cite{Aoki:1982ed} guarantee the unit-residue properties of the diagonal
elements of the renormalized propagator matrix to all orders, in compliance
with the Lehmann-Symanzik-Zimmermann (LSZ) reduction formalism
\cite{Lehmann:1954rq}.
The purpose of this Letter is to construct closed analytic expressions for the
mass counterterms and WFR matrices without resorting to perturbation theory and
to generalize the treatment to unstable fermions.
Strictly speaking, unstable particles are not entitled to appear in asymptotic
states of scattering amplitudes in quantum field theory.
However, in numerous applications of significant phenomenological interest, the
rigorous compliance with this tenet would immediately entail a proliferation of
external legs and bring the evaluation of radiative corrections to a grinding
halt, the more so as almost all the known elementary particles are unstable.

We consider a system of $N$ unstable Dirac fermions in the context of some
general parity-nonconserving renormalizable quantum field theory with
inter-generation mixing, such as the up-type or down-type quarks in the SM.
We start from the bare theory and assume that the mass matrix has already been
diagonalized.
The bare masses $m_i^0$, where $i=1,\ldots,N$ is the generation index and the
superscript 0 labels bare quantities, are real and non-negative to ensure the
reality of the action and the absence of tachyons, respectively.
For the sake of a compact notation, we group the bare quantum fields
$\psi_i^0(x)$ into a column vector $\Psi^0(x)$.
In momentum space, the unrenormalized propagator matrix is defined as
$iP(\slashed{p})=
\int d^4x\,e^{ip\cdot x}\langle0|T[\Psi^0(x)\bar{\Psi}^0(0)]|0\rangle$,
where $T$ is the time-ordered product,
$\bar{\Psi}^0(x)=[\Psi^0(x)]^\dagger\gamma^0$, and a tensorial product
both in the spinor and generation spaces is implied.
Its inverse is built up by the one-particle-irreducible Feynman diagrams
contributing to the transitions $j\to i$ and takes the form
\begin{equation}
P^{-1}(\slashed{p})=\slashed{p}-M^0-\Sigma(\slashed{p}),
\label{eq:bareinv}
\end{equation}
where $M_{ij}^0=m_i^0\delta_{ij}$ and $\Sigma(\slashed{p})$ is the
unrenormalized self-energy matrix.
Here and in the following, repeated indices are not summed over unless
indicated by a summation symbol.
Lorentz covariance entails 
\begin{equation}
\Sigma(\slashed{p})=[\slashed{p}B_+(p^2)+A_+(p^2)]a_+
+(+\leftrightarrow-),
\end{equation}
where $a_\pm=(1\pm\gamma_5)/2$ are the chiral projection operators and
$[A_\pm(p^2)]_{ij}$ and $[B_\pm(p^2)]_{ij}$ are Lorentz invariants.
The latter may be calculated from the bare Lagrangian order by order in
perturbation theory.
However, we refrain from resorting to perturbative expansions 
so as render our results valid to all orders.
Defining
\begin{equation}
S_\pm(p^2)=\mathbbm{1}-B_\pm(p^2),\qquad T_\pm=M^0+A_\pm(p^2),
\label{eq:stbare}
\end{equation}
Eq.~(\ref{eq:bareinv}) becomes
\begin{equation}
P^{-1}(\slashed{p})=[\slashed{p}S_+(p^2)-T_+(p^2)]a_+
+(+\leftrightarrow-).
\label{eq:bareinv1}
\end{equation}
Performing the Dyson resummation \cite{Dyson:1949ha} is equivalent to inverting
Eq.~(\ref{eq:bareinv1}) and yields \cite{Kniehl:2012zb}
\begin{eqnarray}
\lefteqn{P(\slashed{p})}
\label{eq:bare}\\
&=&[\slashed{p}+D_-(p^2)]S_-^{-1}(p^2)[p^2-E_-(p^2)]^{-1}a_+
+(+\leftrightarrow-)
\nonumber\\
&=&a_+[p^2-F_+(p^2)]^{-1}S_+^{-1}(p^2)[\slashed{p}+C_+(p^2)]
+(+\leftrightarrow-),
\nonumber
\end{eqnarray}
with the short-hand notations
\begin{eqnarray}
C_\pm(p^2)&=&T_\mp(p^2)S_\mp^{-1}(p^2),\
D_\pm(p^2)=S_\mp^{-1}(p^2)T_\pm(p^2),
\nonumber\\
E_\pm(p^2)&=&C_\pm(p^2)C_\mp(p^2),\
F_\pm(p^2)=D_\mp(p^2)D_\pm(p^2),
\quad\label{eq:cdbare}
\end{eqnarray}
where $S_\pm^{-1}(p^2)=\sum_{n=0}^\infty B_\pm^n(p^2)$ is a geometric series.

In the following, we shall exploit several times the following theorem for
$n\times n$ matrices $A$ (for a proof, see e.g.\ Ref.~\cite{book}):
\begin{equation}
A(\adj A)=(\adj A)A=(\det A)\mathbbm{1},
\label{eq:adj}
\end{equation}
where $(\adj A)_{ij}=C_{ji}$ with $C_{ij}$ being the cofactor of $A_{ij}$,
i.e.\ $(-1)^{i+j}$ times the determinant of the $(n-1)\times(n-1)$ matrix
obtained by deleting the $i$-th row and the $j$-th column of $A$.
If $\det A\ne0$, then Eq.~(\ref{eq:adj}) implies that
$A^{-1}=(\adj A)/(\det A)$.
Since the four matrices $[p^2-E_\pm(p^2)]$ and $[p^2-F_\pm(p^2)]$ are related
by similarity transformations, their determinants coincide.
Owing to Eq.~(\ref{eq:adj}), the individual propagator parts in
Eq.~(\ref{eq:bare}) thus all have their poles at the same (complex) positions
$p^2=M_i^2$, which are the zeros of any of the secular equations
\cite{Donoghue:1979jq,Gambino:1999ai,Kniehl:2012zb}
\begin{equation}
\det[M_i^2-E_\pm(M_i^2)]=\det[M_i^2-F_\pm(M_i^2)]=0.
\label{eq:zero}
\end{equation}
Here, $M_i$ is the complex pole mass of fermion $i$, which is related to the
real pole mass $m_i$ and total decay width $\Gamma_i$ as
\cite{Smith:1996xz,Kniehl:2008cj}
\begin{equation}
M_i=m_i-i\frac{\Gamma_i}{2}.
\label{eq:real}
\end{equation}
In the pole renormalization scheme, $M_i$ serve as the renormalized masses,
i.e.\ the mass counterterms $\delta M_i$ are fixed by
\begin{equation}
m_i^0=M_i+\delta M_i.
\label{eq:mren}
\end{equation}

We now turn to the WFR.
We first assume that all the fermions are stable, with $\Gamma_i=0$, i.e.\
their mass shells $p^2=m_i^2$ lie below the thresholds of $[A_\pm(p^2)]_{ij}$
and $[B_\pm(p^2)]_{ij}$, where the absorptive parts of the latter vanish.
The WFR is implemented by writing
\begin{equation}
\Psi^0(x)=Z^{\nicefrac{1}{2}}\Psi(x),
\label{eq:wfr}
\end{equation}
where $\Psi(x)$ is the renormalized field multiplet and 
\begin{equation}
Z^{\nicefrac{1}{2}}=Z_+^{\nicefrac{1}{2}}a_++Z_-^{\nicefrac{1}{2}}a_-
\label{eq:wfrc}
\end{equation}
is the WFR matrix. 
Using Eq.~(\ref{eq:wfr}), we may express the renormalized propagator matrix
$i\hat{P}(\slashed{p})=
\int d^4x\,e^{ip\cdot x}\langle0|T[\Psi(x)\bar{\Psi}(0)]|0\rangle$
in terms of the unrenormalized one as
\begin{equation}
\hat{P}(\slashed{p})
=Z^{-\nicefrac{1}{2}}P(\slashed{p})\bar{Z}^{-\nicefrac{1}{2}},
\label{eq:rel}
\end{equation}
where
\begin{equation}
\bar{Z}^{\nicefrac{1}{2}}=\gamma^0Z^{\dagger\nicefrac{1}{2}}\gamma^0
=a_-Z_+^{\dagger\nicefrac{1}{2}}+a_+Z_-^{\dagger\nicefrac{1}{2}}.
\label{eq:wfrd}
\end{equation}
We may absorb the WFR matrices in Eq.~(\ref{eq:rel}) by introducing
renormalized counterparts of $S_\pm$ and $T_\pm$ in Eq.~(\ref{eq:stbare}),
\begin{equation}
\hat{S}_\pm(p^2)
=Z_\pm^{\dagger\nicefrac{1}{2}}S_\pm(p^2)Z_\pm^{\nicefrac{1}{2}},
\quad
\hat{T}_\pm(p^2)
=Z_\mp^{\dagger\nicefrac{1}{2}}T_\pm(p^2)Z_\pm^{\nicefrac{1}{2}}.
\label{eq:st}
\end{equation}
Feeding Eq.~(\ref{eq:st}) into Eq.~(\ref{eq:cdbare}), we are thus led to define
\begin{eqnarray}
\hat{C}_\pm(p^2)&=&\hat{T}_\mp(p^2)\hat{S}_\mp^{-1}(p^2)
=Z_\pm^{\dagger\nicefrac{1}{2}}C_\pm(p^2)Z_\mp^{\dagger-\nicefrac{1}{2}},
\nonumber\\
\hat{D}_\pm(p^2)&=&\hat{S}_\mp^{-1}(p^2)\hat{T}_\pm(p^2)
=Z_\mp^{-\nicefrac{1}{2}}D_\pm(p^2)Z_\pm^{\nicefrac{1}{2}},
\nonumber\\
\hat{E}_\pm(p^2)&=&\hat{C}_\pm(p^2)\hat{C}_\mp(p^2)
=Z_\pm^{\dagger\nicefrac{1}{2}}E_\pm(p^2)Z_\pm^{\dagger-\nicefrac{1}{2}},
\nonumber\\
\hat{F}_\pm(p^2)&=&\hat{D}_\mp(p^2)\hat{D}_\pm(p^2)
=Z_\pm^{-\nicefrac{1}{2}}F_\pm(p^2)Z_\pm^{\nicefrac{1}{2}}.
\label{eq:cd}
\end{eqnarray}
The renormalized counterparts of Eqs.~(\ref{eq:bareinv1}) and (\ref{eq:bare})
then simply emerge by placing carets.
From Eq.~(\ref{eq:cd}), we learn that the matrices $\hat{E}_\pm(p^2)$ and
$E_\pm(p^2)$ ($\hat{F}_\pm(p^2)$ and $F_\pm(p^2)$) are similar, which implies
that their determinants coincide.
Hence, the pole positions $M_i^2$ fixed by Eq.~(\ref{eq:zero}) are not affected
by the WFR \cite{Gambino:1999ai}.

In accordance with the LSZ reduction formalism \cite{Lehmann:1954rq}, we
determine $Z^{\nicefrac{1}{2}}$ by requiring that, if the mass shell of a
fermion is reached, the respective diagonal element of the renormalized
propagator matrix resonates with unit residue, while the other elements stay
finite, i.e.\
\begin{equation}
[\hat{P}(\slashed{p})]_{ij}=
\frac{\delta_{in}\delta_{nj}}{\slashed{p}-M_n}+\mathcal{O}(1),
\label{eq:unit}
\end{equation}
in the limit $p^2\to M_n^2$. 
For $\Gamma_i=0$, this may be achieved by imposing the on-shell WFR conditions
\cite{Aoki:1982ed},
\begin{eqnarray}
[\hat{P}^{-1}(\slashed{p})]_{ij}u(\vec{p},M_j)&=&0,
\label{eq:aoki1}\\
\bar{u}(\vec{p},M_i)[\hat{P}^{-1}(\slashed{p})]_{ij}&=&0,
\label{eq:aoki2}\\
\left\{\frac{1}{\slashed{p}-M_i}[\hat{P}^{-1}(\slashed{p})]_{ii}\right\}
u(\vec{p},M_i)&=&u(\vec{p},M_i),
\label{eq:aoki3}\\
\bar{u}(\vec{p},M_i)
\left\{[\hat{P}^{-1}(\slashed{p})]_{ii}\frac{1}{\slashed{p}-M_i}\right\}
&=&\bar{u}(\vec{p},M_i),
\label{eq:aoki4}
\end{eqnarray}
for all $i,j=1,\ldots,N$, where $u(\vec{p},M_i)$ is a four-component
spinor satisfying the Dirac equation $(\slashed{p}-M_i)u(\vec{p},M_i)=0$ and
$\bar{u}(\vec{p},M_i)=[u(\vec{p},M_i)]^\dagger\gamma^0$.
For $\Gamma_i=0$, an explicit proof that
Eqs.~(\ref{eq:aoki1})--(\ref{eq:aoki4}) entail Eq.~(\ref{eq:unit}) may be found
in Sec.~III of Ref.~\cite{Kniehl:2012zb}.
Equations~(\ref{eq:aoki1})--(\ref{eq:aoki3}) imply that
\begin{eqnarray}
0&=&[\hat{S}_\mp(M_j^2)]_{ij}M_j-[\hat{T}_\pm(M_j^2)]_{ij},
\label{eq:aoki1a}\\
0&=&M_i[\hat{S}_\pm(M_i^2)]_{ij}-[\hat{T}_\pm(M_i^2)]_{ij},
\label{eq:aoki2a}\\
1&=&[\hat{S}_+(M_i^2)]_{ii}+M_i^2\{[\hat{S}_+^\prime(M_i^2)]_{ii}
+[\hat{S}_-^\prime(M_i^2)]_{ii}\}\nonumber\\
&&{}-M_i\{[\hat{T}_+^\prime(M_i^2)]_{ii}+[\hat{T}_-^\prime(M_i^2)]_{ii}\},
\label{eq:aoki3a}
\end{eqnarray}
respectively, while Eq.~(\ref{eq:aoki4}) is redundant.
Equation~(\ref{eq:aoki3}) also implies that
$[\hat{S}_+(M_i^2)]_{ii}=[\hat{S}_-(M_i^2)]_{ii}$, which, however, already
follows from Eqs.~(\ref{eq:aoki1a}) and (\ref{eq:aoki2a}) for $i=j$.

We now solve Eqs.~(\ref{eq:aoki1a})--(\ref{eq:aoki3a}) exactly for $M_i$,
$Z^{\nicefrac{1}{2}}$, and $Z^{\dagger\nicefrac{1}{2}}$, without recourse to
perturbation theory.
Multiplying Eq.~(\ref{eq:aoki1a}) by $[\hat{S}_\mp^{-1}(M_j^2)]_{ki}$ from the
left, summing over $i$, iterating the outcome, and proceeding analogously with
Eq.~(\ref{eq:aoki2a}), we obtain the following eigenvalue equations:
\begin{eqnarray}
{[F_\pm(M_j^2)Z_\pm^{\nicefrac{1}{2}}]}_{ij}&=&
(Z_\pm^{\nicefrac{1}{2}})_{ij}M_j^2,
\nonumber\\
{[Z_\pm^{\dagger\nicefrac{1}{2}}E_\pm(M_i^2)]}_{ij}&=&
M_i^2(Z_\pm^{\dagger\nicefrac{1}{2}})_{ij}.
\label{eq:z}
\end{eqnarray}
With the aid of Eqs.~(\ref{eq:adj}) and (\ref{eq:zero}), we find solutions of
the form
\begin{equation}
(Z_\pm^{\nicefrac{1}{2}})_{ij}=M_{ij}^\pm\lambda_j^\pm,
\qquad
(Z_\pm^{\dagger\nicefrac{1}{2}})_{ij}=\bar{\lambda}_i^\pm\bar{M}_{ij}^\pm,
\label{eq:z2}
\end{equation}
where $\lambda_i^\pm$ and $\bar{\lambda}_i^\pm$ are constants yet to be
determined and
\begin{eqnarray}
M_{ij}^\pm&=&\{\adj[M_j^2-F_\pm(M_j^2)]\}_{ij},
\nonumber\\
\bar{M}_{ij}^\pm&=&\{\adj[M_i^2-E_\pm(M_i^2)]\}_{ij}.
\label{eq:m}
\end{eqnarray}
Substituting Eq.~(\ref{eq:z2}) into Eqs.~(\ref{eq:aoki1a}) and
(\ref{eq:aoki2a}) with $i=j$ and Eq.~(\ref{eq:aoki3a}), we have
\begin{eqnarray}
&&\hspace{-0.7cm}M_i\bar{\lambda}_i^+s_i^+\lambda_i^+
=M_i\bar{\lambda}_i^-s_i^-\lambda_i^-
=\bar{\lambda}_i^-t_i^+\lambda_i^+
=\bar{\lambda}_i^+t_i^-\lambda_i^-,
\label{eq:aokia}\\
&&\hspace{-0.7cm}\bar{\lambda}_i^+s_i^+\lambda_i^+
+M_i^2(\bar{\lambda}_i^+s_i^{+\prime}\lambda_i^+
+\bar{\lambda}_i^-s_i^{-\prime}\lambda_i^-)
\nonumber\\
&&\hspace{-0.7cm}{}-M_i(\bar{\lambda}_i^-t_i^{+\prime}\lambda_i^+
+\bar{\lambda}_i^+t_i^{-\prime}\lambda_i^-)=1,
\label{eq:aokib}
\end{eqnarray}
where
\begin{equation}
s_i^\pm(p^2)=[\bar{M}^\pm S_\pm(p^2)M^\pm]_{ii},\
t_i^\pm(p^2)=[\bar{M}^\mp T_\pm(p^2)M^\pm]_{ii},
\end{equation}
and $p^2=M_i^2$ is implied whenever the arguments are omitted.
From Eq.~(\ref{eq:aokia}), we obtain
\begin{equation}
M_i^2=f_i(M_i^2),
\label{eq:mass}
\end{equation}
where
\begin{equation}
f_i(p^2)=\frac{t_i^+(p^2)t_i^-(p^2)}{s_i^+(p^2)s_i^-(p^2)}.
\label{eq:fi}
\end{equation}
Factoring out $\bar{\lambda}_i^+s_i^+\lambda_i^+$ in Eq.~(\ref{eq:aokib}) and
using Eqs.~(\ref{eq:aokia}), (\ref{eq:mass}), and (\ref{eq:fi}), we find
\begin{equation}
\bar{\lambda}_i^+s_i^+\lambda_i^+[1-f_i^\prime(M_i^2)]=1.
\label{eq:aokic}
\end{equation}
Using Eq.~(\ref{eq:z2}) for $i=j$ and Eq.~(\ref{eq:aokia}), we arrive at
\begin{eqnarray}
(Z_\pm^{\dagger\nicefrac{1}{2}})_{ii}(Z_\pm^{\nicefrac{1}{2}})_{ii}&=&
\frac{\bar{M}_{ii}^\pm M_{ii}^\pm}{s_i^\pm[1-f_i^\prime(M_i^2)]},
\label{eq:zzs}\\
(Z_\mp^{\dagger\nicefrac{1}{2}})_{ii}(Z_\pm^{\nicefrac{1}{2}})_{ii}&=&
\frac{M_i\bar{M}_{ii}^\mp M_{ii}^\pm}{t_i^\pm[1-f_i^\prime(M_i^2)]}.
\label{eq:zzt}
\end{eqnarray}
The nondiagonal entities are then fixed by Eq.~(\ref{eq:z2}) to be
\begin{equation}
(Z_\pm^{\nicefrac{1}{2}})_{ij}=\frac{M_{ij}^\pm}{M_{jj}^\pm}
(Z_\pm^{\nicefrac{1}{2}})_{jj},\
(Z_\pm^{\dagger\nicefrac{1}{2}})_{ij}=(Z_\pm^{\dagger\nicefrac{1}{2}})_{ii}
\frac{\bar{M}_{ij}^\pm}{\bar{M}_{ii}^\pm}.
\label{eq:z3}
\end{equation}

Owing to our assumption $\Gamma_i=0$, the bare propagator matrix satisfies the
pseudo-Hermiticity condition
$\gamma^0P^\dagger(\slashed{p})\gamma^0=P(\slashed{p})$ \cite{Aoki:1982ed},
which implies that $A_\pm^\dagger(p^2)=A_\mp(p^2)$ and
$B_\pm^\dagger(p^2)=B_\pm(p^2)$ \cite{Kniehl:1996bd,Espriu:2002xv}.
Hence, we have $F_\pm^\dagger(p^2)=E_\pm(p^2)$,
$(M^\pm)^\dagger=\bar{M}^\pm$, $[s_i^\pm(p^2)]^*=s_i^\pm(p^2)$,
$[t_i^\pm(p^2)]^*=t_i^\mp(p^2)$, and $[f_i(p^2)]^*=f_i(p^2)$.
Consequently, the r.h.s.\ of Eq.~(\ref{eq:zzs}) is real, as required by the
l.h.s.\ being $|(Z_\pm^{\nicefrac{1}{2}})_{ii}|^2$, and complex conjugation of
Eq.~(\ref{eq:zzt}) entails a flip of the alternating-sign labels on both sides.
Furthermore, Eqs.~(\ref{eq:mass}) and (\ref{eq:zzs})--(\ref{eq:z3}) are
consistent with each other.
For each value of $i$, Eqs.~(\ref{eq:zzs}) and (\ref{eq:zzt}) provide four real
equations for the four real unknowns $\re(Z_\pm^{\nicefrac{1}{2}})_{ii}$ and
$\im(Z_\pm^{\nicefrac{1}{2}})_{ii}$.
However, one of these equations is redundant due to Eq.~(\ref{eq:mass}).
We may exhaust this residual freedom by choosing e.g.\
$(Z_+^{\dagger\nicefrac{1}{2}})_{ii}=(Z_+^{\nicefrac{1}{2}})_{ii}$, as in
Ref.~\cite{Kniehl:1996bd}.
This freedom does not affect Eq.~(\ref{eq:unit}).
In fact, Eqs.~(\ref{eq:mass}) and (\ref{eq:zzs})--(\ref{eq:z3}) are valid to
all orders. 
At one loop, they agree with Eqs.~(3.13) and (3.15)--(3.17) in
Ref.~\cite{Kniehl:1996bd} and with Eqs.~(3.3), (3.4), (4.3), and (4.4) in
Ref.~\cite{Espriu:2002xv}.
At two loops, Eq.~(\ref{eq:mass}) coincides with Eq.~(23) in
Ref.~\cite{Kniehl:2012zb}, which was found there by directly solving
Eq.~(\ref{eq:zero}).

We now allow for some or all of the fermions to be unstable, releasing $M_i$ to
complex values.
This immediately leads to contradictions because the r.h.s.\ of
Eq.~(\ref{eq:zzs}) is no longer real and that of Eq.~(\ref{eq:zzt}) no longer
flips the alternating-sign labels upon complex conjugation, while the l.h.s.'s
still possess these properties.
This problem may be cured by allowing the WFR matrices of the in and out states
to bifurcate when $\Gamma_i$ increase to assume their physical values, as was
already noticed in the pioneering one-loop analysis of
Ref.~\cite{Espriu:2002xv}.
This amounts to abandoning the first equality in Eq.~(\ref{eq:wfrd}) and 
replacing everywhere $Z_\pm^\dagger$ by $\bar{Z}_\pm$, say.
Since the above manipulations of Eq.~(\ref{eq:rel}) actually never rely on the
relationship between Eqs.~(\ref{eq:wfrc}) and (\ref{eq:wfrd}), the derivation
of Eqs.~(\ref{eq:mass}) and (\ref{eq:zzs})--(\ref{eq:z3}) carries over without
further ado, and so does the proof \cite{Kniehl:2012zb} that
Eqs.~(\ref{eq:aoki1})--(\ref{eq:aoki4}) guarantee Eq.~(\ref{eq:unit}). 
For each value of $i$, Eqs.~(\ref{eq:zzs}) and (\ref{eq:zzt}) now provide four
complex equations for the four complex unknowns
$(Z_\pm^{\nicefrac{1}{2}})_{ii}$ and
$(\bar{Z}_\pm^{\nicefrac{1}{2}})_{ii}$.
However, one of these equations is redundant, and we may express any three of
the unknowns in terms of the fourth one.
We may exploit this liberty e.g.\ by identifying
$(\bar{Z}_+^{\nicefrac{1}{2}})_{ii}=(Z_+^{\nicefrac{1}{2}})_{ii}$.
Again, this does not affect Eq.~(\ref{eq:unit}).
From Eqs.~(\ref{eq:mren}) and (\ref{eq:mass}), we obtain the all-order mass
counterterms as
\begin{equation}
\delta M_i=m_i^0-\sqrt{f_i(M_i^2)}.
\label{eq:mct}
\end{equation} 
Using also Eq.~(\ref{eq:real}), we have
\begin{eqnarray}
m_i&=&\re\sqrt{f_i(M_i^2)}=m_i^0-\re\delta M_i,
\label{eq:realmass}\\
-\frac{\Gamma_i}{2}&=&\im\sqrt{f_i(M_i^2)}=-\im\delta M_i.
\label{eq:width}
\end{eqnarray}

Expanding the building blocks of Eqs.~(\ref{eq:mass}) and
(\ref{eq:zzs})--(\ref{eq:z3}) through $\mathcal{O}(\alpha^2)$, we find
\begin{eqnarray}
f_i(p^2)&=&
\frac{[T_+(p^2)]_{ii}[T_-(p^2)]_{ii}}{[S_+(p^2)]_{ii}[S_-(p^2)]_{ii}}
+m_i^0[\tau_i^+(p^2)+\tau_i^-(p^2)]
\nonumber\\
&&{}-(m_i^0)^2[\sigma_i^+(p^2)+\sigma_i^-(p^2)]
+\mathcal{O}(\alpha^3),
\nonumber\\
\frac{s_i^\pm(p^2)}{\bar{M}_{ii}^\pm M_{ii}^\pm}&=&
[S_\pm(p^2)]_{ii}+\sigma_i^\pm(p^2)
+\mathcal{O}(\alpha^3),
\nonumber\\
\frac{t_i^\pm(p^2)}{\bar{M}_{ii}^\mp M_{ii}^\pm}&=&
[T_\pm(p^2)]_{ii}+\tau_i^\pm(p^2)
+\mathcal{O}(\alpha^3),
\nonumber\\
\frac{M_{ji}^\pm}{M_{ii}^\pm}&=&
f_{iji}^\pm(1+f_{ijj}^\pm)+\sum_{i\ne k\ne j}f_{ijk}^\pm f_{iki}^\pm
+\mathcal{O}(\alpha^3),
\nonumber\\
\frac{\bar{M}_{ij}^\pm}{\bar{M}_{ii}^\pm}&=&
e_{iij}^\pm(1+e_{ijj}^\pm)+\sum_{i\ne k\ne j}e_{iik}^\pm f_{ikj}^\pm
+\mathcal{O}(\alpha^3),
\end{eqnarray}
for $j\ne i$, where
\begin{eqnarray}
\sigma_i^\pm(p^2)&=&
\sum_{j\ne i}\{e_{iij}^\pm f_{iji}^\pm-e_{iij}^\pm[B_\pm(p^2)]_{ji}
-[B_\pm(p^2)]_{ij}f_{iji}^\pm\},
\nonumber\\
\tau_i^\pm(p^2)&=&
\sum_{j\ne i}\{e_{iij}^\mp m_j^0f_{iji}^\pm+e_{iij}^\mp[A_\pm(p^2)]_{ji}
\nonumber\\
&&{}+[A_\pm(p^2)]_{ij}f_{iji}^\pm\},
\nonumber\\
f_{ijk}^\pm&=&
\frac{F_\pm(M_i^2)]_{jk}-M_j^2\delta_{jk}}{M_i^2-M_j^2}\quad(j\ne i),
\nonumber\\
e_{ijk}^\pm&=&
\frac{E_\pm(M_i^2)]_{jk}-\delta_{jk}M_k^2}{M_i^2-M_k^2}\quad(k\ne i).
\end{eqnarray}

We now consider the special case of a single unstable fermion,
in which Eqs.~(\ref{eq:mass}) and (\ref{eq:zzs})--(\ref{eq:z3}) collapse and
uniquely determine the renormalized propagator to be
\begin{eqnarray}
\hat{P}(\slashed{p})&=&\left[\slashed{p}
+M\frac{S_+(M^2)}{S_+(p^2)}\,\frac{T_-(p^2)}{T_-(M^2)}\right]
\frac{S_-(M^2)}{S_-(p^2)}
\nonumber\\
&&{}\times\frac{1-f^\prime(M^2)}{p^2-f(p^2)}a_++(+\leftrightarrow-),
\label{eq:single}
\end{eqnarray}
where $f(p^2)=T_+(p^2)T_-(p^2)/[S_+(p^2)S_-(p^2)]$ and $M^2=f(M^2)$.
Evidently, Eq.~(\ref{eq:single}) has unit residue at the physical pole
$\slashed{p}=M$.
We note that Eq.~(\ref{eq:single}) slightly differs from Eq.~(36) in
Ref.~\cite{Kniehl:2008cj}, where a renormalization scheme without WFR
bifurcation was employed.

In summary, we renormalized the propagator matrix of a mixed system of Dirac
fermions in a general parity-nonconserving quantum field theory adopting the
pole scheme, in which the pole masses $M_i$ serve as the renormalized masses
and the WFR matrices $Z_\pm^{\nicefrac{1}{2}}$ and
$Z_\pm^{\dagger\nicefrac{1}{2}}$ are adjusted in compliance with the LSZ
reduction formalism \cite{Lehmann:1954rq}.
We derived closed analytic expressions for the renormalization constants in
terms of the scalar, pseudoscalar, vector, and pseudovector parts of the
unrenormalized self-energy matrix.
These are valid to all orders and reproduce the results available in the
literature, for $M_i$ at one \cite{Kniehl:1996bd,Espriu:2002xv} and two loops
\cite{Kniehl:2012zb} and for $Z_\pm^{\nicefrac{1}{2}}$ and
$Z_\pm^{\dagger\nicefrac{1}{2}}$ at one loop
\cite{Kniehl:1996bd,Espriu:2002xv}.
We identified residual freedom in the determination of
$Z_\pm^{\nicefrac{1}{2}}$ and $Z_\pm^{\dagger\nicefrac{1}{2}}$ and proposed
an additional renormalization condition to exhaust it.
We then explained how our results carry over from stable fermions to unstable
ones.
In the latter case, we encountered WFR bifurcation, i.e.\ the departure from
the first equality in Eq.~(\ref{eq:wfrd}), confirming the findings of
Ref.~\cite{Espriu:2002xv} at one loop.

Apart from being conceptually interesting in their own right, our results have
a number of important phenomenological applications, of which we mention but
three below.
First, in the perturbative treatment of a specific particle scattering or decay
process involving unstable fermions, such as top-quark production and decay,
Eqs.~(\ref{eq:mass}) and (\ref{eq:zzs})--(\ref{eq:z3}) may be readily employed,
after expansion through the considered order and truncation of terms beyond
that order.
Second, the total decay width $\Gamma_i$, e.g.\ that of the top quark, may be
conveniently evaluated through any order from $[A_\pm(p^2)]_{ij}$ and
$[B_\pm(p^2)]_{ij}$ by solving Eq.~(\ref{eq:width}) iteratively.
Third, Eqs.~(\ref{eq:mren}) and (\ref{eq:mct}) may be used to switch from the
pole scheme adopted here to any other scheme of mass renormalization, as long
as the method of regularization is maintained, exploiting the scheme
independence of $m_i^0$.
In this way, the $\overline{\mathrm{MS}}$ \cite{Bardeen:1978yd} definition of
mass may be naturally extended from QCD to the EW sector, as
\begin{equation}
\bar{m}_i=m_i+(\re\delta M_i)_{\overline{\mathrm{MS}}},
\end{equation}
where $(\re\delta M_i)_{\overline{\mathrm{MS}}}$ is the UV-finite remainder of
$\re\delta M_i$ after $\overline{\mathrm{MS}}$ subtraction of the poles in
$\varepsilon=2-d/2$ at renormalization scale $\mu$, where $d$ is the
dimensionality of space time in dimensional regularization
\cite{Bollini:1972ui}.
In spontaneously broken gauge theories, such as the SM, it is necessary to
include the tadpole contributions in
$(\re\delta M_i)_\mathrm{\overline{\mathrm{MS}}}$ in order for $\bar{m}_i$
to be gauge independent \cite{Hempfling:1994ar}.
In the case of the top quark, the accumulated QCD contribution to
$(\re\delta M_t)_\mathrm{\overline{\mathrm{MS}}}$ from orders
$\mathcal{O}(\alpha_s^n)$ with $n=1,2,3$, which renders $\bar{m}_t$ at
$\mu=m_t$ approximately 10~GeV smaller than $m_t$, happens to be almost
perfectly compensated by the EW contribution from orders
$\mathcal{O}(\alpha\alpha_s^n)$ with $n=0,1$ for $M_H\approx126$~GeV
\cite{Jegerlehner:2012kn}.

We are indebted to Alberto Sirlin for numerous beneficial discussions.
This research was supported in part by 
DFG Grant No.\ SFB 676 and by 
NSF Grant No.\ PHY11-25915.

\end{document}